\documentclass{entcs} 
\usepackage{prentcmacro}
\usepackage{graphicx}
\usepackage{amssymb}
\usepackage{amsfonts}
\usepackage{latexsym}
\usepackage{epsfig}
\usepackage{verbatim}

\input{diagrams.sty}
\newarrow{Toto}<--->
\newarrow{Is}=====
\newarrow{Dotsto}....>
\newarrow{Dots}.....

\newcommand{\bit}{\begin{itemize}}
\newcommand{\eit}{\end{itemize}\par\noindent}
\newcommand{\ben}{\begin{enumerate}}
\newcommand{\een}{\end{enumerate}\par\noindent}
\newcommand{\beq}{\begin{equation}}
\newcommand{\eeq}{\end{equation}\par\noindent}
\newcommand{\beqa}{\begin{eqnarray*}}
\newcommand{\eeqa}{\end{eqnarray*}\par\noindent}
\newcommand{\beqn}{\begin{eqnarray}}
\newcommand{\eeqn}{\end{eqnarray}\par\noindent}

\newcommand{\bpf}{\noindent {\bf Proof:} }
\def\endproof{\hfill$\Box$}

\def\PP{{\rm P}}

\def\CC{{\bf C}}
\def\II{{\rm I}}

\newcommand{\uu}{\ulcorner}
\newcommand{\uuu}{\urcorner}

\newcommand{\dist}{\mbox{\sf\footnotesize DIST}}

\def\lastname{Coecke}
\begin{document}
\begin{frontmatter}
  \title{De-linearizing Linearity: Projective Quantum Axiomatics from
Strong Compact Closure} \author{Bob Coecke\thanksref{ALL}\thanksref{myemail}}
  \address{Oxford University Computing Laboratory,\\
    Wolfson Building, Parks Road,\\ OX1 3QD Oxford, UK.}  
    \thanks[ALL]{The research which led to these results was
supported by the EPSRC grant EP/C500032/1 on High-Level Methods in
Quantum Computation and Quantum Information.} 
    \thanks[myemail]{Email:
    \href{mailto:coecke@comlab.ox.ac.uk} {\texttt{\normalshape
        coecke@comlab.ox.ac.uk}}} 
\begin{abstract} 
Elaborating on our joint work with Abramsky in
\cite{AC1,AC1.5} we further unravel the linear structure of Hilbert
spaces into several constituents.   Some prove to be very crucial for particular features of
quantum theory while others obstruct the passage to a formalism  which is
not saturated with physically  insignificant global phases.

First we show that the bulk of the required linear structure is purely multiplicative, and arises from the strongly compact closed tensor which, besides providing a variety of notions
such as scalars, trace,  unitarity, self-adjointness and bipartite
projectors \cite{AC1,AC1.5},   also provides \em Hilbert-Schmidt norm\em,
\em Hilbert-Schmidt inner-product\em, and in particular, the \em preparation-state agreement
axiom \em which enables the passage from a formalism of the vector space kind to a rather
projective one, as it was intended in the (in)famous Birkhoff \& von
Neumann paper
\cite{BvN}.  

Next we consider additive types which distribute over the tensor,
from which measurements can be build, and the correctness proofs
of the protocols discussed in \cite{AC1} carry over to the resulting weaker setting. A full
probabilistic calculus is obtained when the trace is moreover \em
linear \em and satisfies the \em diagonal axiom\em, which
brings us to a second main result, characterization of the necessary and
sufficient additive structure of a both qualitatively and
quantitatively effective categorical quantum formalism without
redundant global phases. Along the way we show that if in a category a (additive) monoidal tensor distributes over a strongly compact closed tensor, then this category is always enriched in commutative monoids. 
\end{abstract}
\begin{keyword}
Strong compact closure, quantum mechanics, global phases, projective geometry, categorical trace, quantum logic.
\end{keyword}
\end{frontmatter}

\section{Introduction}

The formalism  of the most successful physical theory of the previous
century has many redundant and operationally  insignificant ingredients
e.g.~the redundancy of \em global phases\em.    Its creator himself, John von
Neumann  \cite{vN}, was very aware of this fact \cite{Redei}.   The key
insight of the (in)famous 1936 Birkhoff-von Neumann paper entitled
``The Logic of Quantum Mechanics'' \cite{BvN} is that when eliminating redundant
global scalars one passes from a vector space to a \em projective
space\em.  Such a projective space has a non-distributive lattice of
subspaces and hence the deducted natural level of abstraction was a
lattice-theoretic one, but after seven decades there is still no
satisfactory abstract counterpart to the role which the tensor product
plays in von Neumann's Hilbert space formalism. Also, the world of lattices is
insufficiently comprehensive to give any explicit account on
probabilities, which are traditionally left implicit by relying on
Gleason's theorem \cite{Gleason} e.g.~Piron's book \cite{Piron2}. As discussed in \cite{AC1},
another shortcoming of von Neumann's formalism is the total lack of
\em types \em reflecting kinds e.g.~$f:{\cal H}\to{\cal H}$ can be
reversible dynamics, a measurement, either desctructive or
non-destructive, or a mixed state.

Typing and finding an appropriate abstraction of the quantum formalism
was  (re-)addressed by Abramsky and myself in
\cite{AC1} where we recast the formalism of quantum mechanics in purely
category-theoretic terms. We considered
\em strongly compact closed categories \em with \em biproducts \em and
we showed that all the Hilbert space machinery necessary for quantum
mechanics arises in that setting, but now equipped with appropriate
types and high-level tools for reasoning about entanglement --- following the tradition of
\em linear logic \em we will refer to the strongly compact closed
structure as the
\em multiplicative \em part of the structure and to the \em biproducts
\em as the
\em additive \em part of the structure. The abstract counterpart to the
Hilbert space tensor product is now  a structural primitive from which,
surprisingly,  most of the required ingredients for a quantum
formalism can be derived \cite{AC1,AC1.5}.  Hence we postulated the axioms of (finitary
dimensional) quantum mechanics in terms of strong compact closure, and
biproducts, and it turned out that many non-trivial results obtained
within von Neumann's formalism such as quantum teleportation,
logic-gate teleportation and entanglement swapping become almost
trivial in the abstract setting. Moreover, the abstract setting is far
more \em expressive \em and is \em explicitly operational \em (in the
compositional sense), and of course, admits a lot more \em axiomatic
freedom\em, and, last but definitely not least, turns out to still be
a \em quantitative \em setting.   But unfortunately the biproduct
structure comes together with  redundant global phases and also with
semi-additive enrichment,\footnote{Following
Selinger \cite{Selinger1,Selinger2} semi-additive enrichment does
admit a probabilistic interpretation when considering density matrices, but these
only arise from an irriversible construction on the initial biproduct category  cf.~\cite{Selinger2} and Definition
\ref{def:QL} in this paper.} in laymen's terms, a vector space like
calculus which excludes anything of the projective kind which is  non-trivial.  Biproducts as in \cite{AC1}\footnote{By this we mean biproducts as the type for superposition e.g.~defining a qubit as $Q\simeq \II\oplus \II$. In \cite{Selinger1} biproducts do not play this role, they encode classical controle.  In \cite{Selinger2} there are two levels of biproducts, one which generates superposition but of which the pairing operation gets erased in more or less the same manner as we do it in the construction in Def.~\ref{def:QL} of this paper --- see also the paragraph `parallel work' below.} and in particular the pairing operation of the product structure also cause a collapse of the classical information
flow onto the superposition structure, due to which the physically and syntactically
different entities `classical bit equipped with probability weights'
and `qubit' become categorically isomorphic.

The goal of this paper is to address these problems of the additive
part of the structure by reconsidering von Neumann's initial concern
which led to quantum logic, but this time not with Birkhoff but with
category theory as a close friend. While in
\cite{BvN}, starting from a single  Hilbert \em space\em, one first
eliminates global scalars and then aims at finding the appropriate
abstraction, i.e.~diagramatically,

{\small
\begin{diagram}
\hspace{-2.5cm}\mbox{\bf Hilbert space } {\cal H}&&\\
\dTo^{\hspace{-2.5cm}\mbox{\sf kill redundant global scalars}} &
\rdTo^{\sf Birkhoff\mbox{\rm -}von\ Neumann\
\mbox{\cite{BvN}}}&\\
\begin{array}{l}
\hspace{-3.5cm}\mbox{\bf lattice of subspaces }\mathbb{L}({\cal H})
\end{array} &
\rTo_{\hspace{0.1cm}\mbox{\sf go\ abstract}\hspace{1.1cm}} & {\bf \
abstract\ lattices}\hspace{-2.5cm}
\end{diagram} }

\vspace{2mm}\noindent we will start from the whole \em category \em of
finite dimensional Hilbert spaces and linear maps {\bf FdHilb}, with
strongly compact closed categories with `some additive structure' as
its appropriate abstraction, and then study the abstract counterpart to
`elimination of redundant global scalars', i.e.~diagramatically,

{\small
\begin{diagram}
\hspace{-1.5cm}\mbox{\bf FdHilb} &
\rTo^{\hspace{0.7cm}\mbox{\sf go\ abstract}\hspace{0.3cm}} &
\mbox{\bf `vectorial' strong compact
closure}\hspace{-5.3cm}&\hspace{3.6cm}
\\   &
\rdTo_{\sf our\ approach} &
\dTo_{\mbox{\sf \ kill redundant global scalars}\hspace{-2.5cm}}\\ &&
\ \,\mbox{\bf `projective' strong compact closure}\!\!\!\hspace{-5.55cm}
\end{diagram} }

\vspace{2mm}\noindent Since the bulk of the required linear structure
is already present in the strong compact closed structure, there is no
need for commitment to the highly demanding biproduct structure, and
we expose the
\em necessary \em and \em sufficient \em additive structure required
for an effective categorical quantum formalism which includes a
probability calculus, but excludes global phases.   
Abstract counterparts to `eliminating global phases' and `absence of global phases' are introduced in Sections \ref{sec:globphasandHS} and \ref{sec:elliminateph}.
In Sections 
\ref{sec:orthostructure},
\ref{sec:OBstructure}
we study the qualitative and the quantitative structural requirements on the additive component of a categorical quantum formalism, respectively referred to as an \em ortho-structure \em and an \em ortho-Bornian structure\em.
In Section \ref{sec:categoricalsem} we readdress the categorical semantics of 
\cite{AC1} and deal with the above mentioned problem of 
the collapse of the classical information flow onto the superposition structure in two possible ways.
An important physically significant  feature of dumping
biproducts is that the dominant role of the scalars vanishes ---
cf.~in the case of biproducts all finitary morphisms arise as matrices
in the semiring of scalars. The resulting sole significance of a
scalar is that of a probability weight e.g.~there is no connection
anymore with the relative phases responsible for interference
phenomena. We discuss this issue briefly in Section \ref{sec:weight}.

\paragraph{Proofs, details, discussion and some more results.} These can be
found in \cite{DLL} which is an extended version of this paper.
Additional sections includes %:
%\ben\item A discussion of different notions of positivity for scalars and corresponding existence of square-roots in the context of the construction which eliminates global phases (Def.~\ref{def:QL}) and the preparation-state agreement axiom which guarantees absence ofglobal phases (Def.~\ref{def:preparation-state}); \item 
a construction which adds abstract global phases and hence
provides a (partial) converse to Def.~\ref{def:QL}; this yields an
abstract equivalence which resembles the fundamental theorem of
projective geometry relating projective spaces and vector spaces.
%;\item Several open questions and possible elaborations.\een

\paragraph{Other work.} The aim of this paper and the conception of the
utterance `quantum logic' is different from the work by Abramsky and
Duncan in \cite{AD}. Their aim is to find a
geometric model and syntax for automated reasoning within  our
categorical formalism of \cite{AC1,AC1.5} in the spirit of the \em
proof-net \em calculus for linear logic, anticipating on the fact that
many quantum protocols such as quantum teleportation have an
underlying diagrammatic interpretation in terms of the \em quantum
information-flow\em, introduced in  \cite{Coe} and abstractly axiomatized 
by Abramsky and myself 

\paragraph{Parallel work.} Selinger's latest  \cite{Selinger2} and this
paper --- which were simultaneously and independently written --- have
a non-empty intersection.  Our {\it WProj}-construction for strongly
compact closed categories coincides with Selinger's canonical
embedding of a strongly compact closed category ${\bf C}$ in its
category of \em completely positive maps
\em ${\bf CPM}({\bf C})$.   Also in  \cite{Selinger2}, Selinger proposes a graphical
language for strong compact closure for which he proved completeness
for equational reasoning ---  we have been using a similar language in
a  more informal manner \cite{AC1,AC1.5} and continue(d) to do so in
this paper. We also mention the independent work by Baez \cite{Baez}
which relates to the developments in \cite{AC1,AC1.5} and by Kauffman
\cite{Kauffman} which relates to those in \cite{Coe}. 

\paragraph{Subsequent work.} In \cite{BobDusko,BobEric} we take a very different approach than the one proposed in the second part of this paper.  Rather than relying on additive types for describing quantum measurements we abstract over classical data and define quantum measurements purely multiplicatively, by considering self-adjoint Eilenberg-Moore coalgebras for comonads induced by a special kind of internal comonoid. Via Selinger's construction  \cite{Selinger2} we were then able to build a \em decoherence \em morphism which takes into account the informatic irreversibility, exactly what we which to accommodate in this  paper by relaxing the additive structure.  Moreover, this approach extends to POVMs via an abstract variant of Naimark's theorem. It remains to be seen how that approach relates to the results presented here, but it's fair to say that the results in \cite{BobEric,BobDusko} seem at the moment more compelling than those presented in the second part of this paper.  In recent work \cite{CoeSel} we also provide an axiomatic characterization of Selinger's construction, which turns out to be strongly intertwined with the preparation-state agreement axiom that we introduce in the first part of this paper.

\paragraph{Acknowledgements.}  This research was partly encouraged by
concerns and questions  raised by Peter Selinger, Alex Simpson and Frank Valckenborgh on
the presentation of abstract quantum mechanics in \cite{AC1}. Also
feedback by  Vincent Danos, Ross Duncan, Peter Hines, Terry Rudolph,
Phil Scott and in particular comments by Samson Abramsky, Rick Blute
and Peter Selinger on earlier versions were very useful. We used Paul Taylor's 2004 package for commutative diagrams.

\section{Some observations on strong compact
closure}\label{sec:globphasandHS}

Recall that a \em strongly compact closed category \em (SCCC)
\cite{AC1.5} is a symmetric monoidal category (SMC) \cite{MacLane},
hence with unit $\II$, natural isomorphisms
$\lambda_A : A \simeq {\rm I}\otimes A$ and $\rho_A: A \simeq
A\otimes{\rm I}$, associativity $\alpha_{A,B,C}:A\otimes(B\otimes
C)\simeq (A\otimes B)\otimes C$ and symmetry $\sigma_{A,B}:A\otimes
B\simeq B\otimes A$, and, with
\bit
\item an involution
$A\mapsto A^*$ on objects called \emph{dual},
\item a contravariant identity-on-objects monoidal involution
$f\mapsto f^\dagger$ on morphisms  called \emph{adjoint}, and,
\item for each object a distinct morphism
$\eta_A:\II\to A^*\otimes A$ called \emph{unit},
\eit which satisfy
\beq\label{eq:SCCC}
\lambda^\dagger_A\circ (\eta^\dagger_{A^*}\otimes
1_A)\circ(1_A\otimes\eta_A)\circ\rho_A=1_A
\eeq and the coherence condition $\eta_{A^*}=
\sigma_{A^*\!,A}\circ\eta_A$, and all natural isomorphisms $\chi$
of the symmetric monoidal structure should satisfy
$\chi^{-1}=\chi^\dagger$, that is, they are \em unitary\em.  Every
SCCC is also a \em compact closed category \em (CCC)\footnote{It
should be clear to the reader that in the context of this paper a \underline{c}ompact
\underline{c}losed
\underline{c}ategory cannot be confused with  a \underline{c}artesian
\underline{c}losed \underline{c}ategory.} \cite{KellyLaplaza} and we
recall that a CCC is a
$*$-autonomous category
\cite{Barr} with a self-dual tensor i.e.~with natural isomorphisms
$u_{A,B}:(A\otimes B)^*\simeq A^*\otimes B^*$ and $u_{\II} :
\II^* \simeq\II$.  For an SCCC we will assume that $u_{\II}$ is also
unitary and that $u_{A,B}$ is strict.  As shown in \cite{AC1.5} the
adjoint of an SCCC decomposes as
\[
f^\dagger=(f^*)_*=(f_*)^*
\]
where both 
$(-)^*$ and
$(-)_*$ are involutive, respectively  contravariant and covariant, and
have
$A\mapsto A^*$ as action on objects.
We will be using two distinct unfoldings of the \em name
\em
$\uu f\uuu:\II\to A^*\!\otimes B$ of a morphism $f:A\to B$, either the
usual definition, or, the \em absorption lemma \em in
\cite{AC1} (Lemma 3.7), respectively,
\beq\label{eq:names} (1_{A^*}\otimes f)\circ\eta_A\ \,\stackrel{\bf
Defn\!.}{=:}\,\
\uu f\uuu\, \stackrel{\bf Lemma}{:=}\, (f^*\otimes 1_B)\circ\eta_B\,.
\eeq   Still following \cite{AC1.5} each morphism also defines a \em
bipartite projector \em
\[ {\rm P}_f:=\uu f\uuu\circ(\uu f\uuu)^\dagger:A^*\otimes B\to
A^*\otimes B\,.
\]  In any SMC $\CC$ there exists a commutative monoid of \em
scalars\em, namely $\CC(\II,\II)$ the endomorphism monoid of the
tensor unit
\cite{KellyLaplaza}. As in \cite{AC1,AC1.5} we define \em scalar
multiplication \em by setting
\[
s\bullet f:=\lambda_B^{-1}\circ (s\otimes f)\circ\lambda_A:A\to B\,.
\]

\begin{lemma}\label{lm:distscalovtensor} Let $f$ and $g$ be a
morphisms and
$s,t$ scalars in an SMC, then
\[  
\!\!\!\!\!\!(s\bullet f)\circ (t\bullet g)=(s\circ t)\bullet(f\circ g)
\ \ {\rm and}\ \ 
(s\bullet f)\otimes (t\bullet g)=(s\circ
t)\bullet(f\otimes g)\,.\ \
\]
\end{lemma}

Each complex number can be written as $r\cdot e^{i\theta}$ with
$r\in\mathbb{R}$ and $\theta\in[0,2\pi[$ to which we respectively
refer as  the \em amplitude \em and the
\em phase\em.   Quantum theory dictates that the states of quantum
systems are represented as one-dimensional subspaces of a Hilbert
space, that is, (non-zero) vectors in a Hilbert space \em up to a \em
(non-zero)
\em scalar multiple\em. Hence when specifying operations on quantum
systems we need only to express to which vector a vector is mapped
\em up to a \em (non-zero) \em scalar multiple\em. Hence {\bf FdHilb}
is saturated with \em global scalars \em which are superfluous for
quantum theory. If we eliminate these, then, since states are also
encoded as morphisms, we also eliminate the redundancy in their
description. We would moreover like to eliminate these global scalars
using a procedure which applies to any SCCC. But in fact we only want
to eliminate \em global phases\em, since, as shown in \cite{AC1},
\em global amplitudes \em allow us to encode probability weights, and
are crucially intertwined with the abstract inner-product via the
abstract \em Born rule\em.  In the case of {\bf FdHilb}, if
$f=e^{i\theta}\cdot g$ with $\theta\in[0,2\pi[$ for $f,g:{\cal
H}_1\to{\cal H}_2$ then
\[ f\otimes f^\dagger = e^{i\theta}\!\!\cdot g\otimes
(e^{i\theta}\!\!\cdot g)^\dagger=e^{i\theta}\!\!\cdot g\otimes
e^{-i\theta}\!\!\cdot g^\dagger=g\otimes g^\dagger\,.
\]   The following lemma indicates that the passage $f\mapsto f\otimes
f^\dagger$ causes also abstract global phases to vanish in some sort of
similar manner.

\begin{proposition}\label{Pr:phase1} For $f$ and $g$  morphisms and
$s,t$ scalars in an SCCC,
\[ s\bullet f=t\bullet g\ ,\ s\circ s^\dagger=t\circ t^\dagger=1_\II\
\ \
\Longrightarrow\ \ \ f\otimes f^\dagger\!= g\otimes g^\dagger.
\]
\end{proposition}

Observing that $1_\II^{-1}\!=1_\II$ it actually suffices to assume
existence of a scalar $x$ such that
$s\circ s^\dagger=t\circ t^\dagger=x^{-1}$. But the real surprise is
the fact that there exists a converse to Proposition \ref{Pr:phase1}.
It is moreover a stronger result in the sense that it extends beyond
cases where there exists an inverse to
$s\circ s^\dagger=t\circ t^\dagger$. This shows that abstract removal
of global phases is truly genuine and not merely generalization by
analogy.

\begin{proposition}\label{Pr:phase2} For $f$ and $g$  morphisms in an
SCCC with scalars $S$,
\[  f\otimes f^\dagger\!=g\otimes g^\dagger\ \ \ \Longrightarrow\ \ \
\exists s,t\in S\,:\ s\bullet f=t\bullet g\ ,\ s\circ s^\dagger=t\circ
t^\dagger.
\] 
In particular can we set\footnote{By symmetry we could also set $s:=(\uu f\uuu)^\dagger\circ\uu g\uuu$ and $t:=(\uu g\uuu)^\dagger\circ\uu g\uuu$.}
\beq\label{eq:Hsscalers} s:=(\uu f\uuu)^\dagger\circ\uu f\uuu
\qquad{\rm and}\qquad t:=(\uu g\uuu)^\dagger\circ\uu f\uuu.
\eeq
\end{proposition}

We prove Proposition \ref{Pr:phase2} using pictures.  We represent
units by triangles and their adjoints by the same triangle but
depicted upside down where we take a from bottom to top reading
convention. Other morphisms are depicted by square boxes.  E.g.~the
scalar $s:=(\uu f\uuu)^\dagger\circ\uu f\uuu$ is depicted as

\vspace{3mm}\noindent{
\begin{minipage}[b]{1\linewidth}
\centering{\epsfig{figure=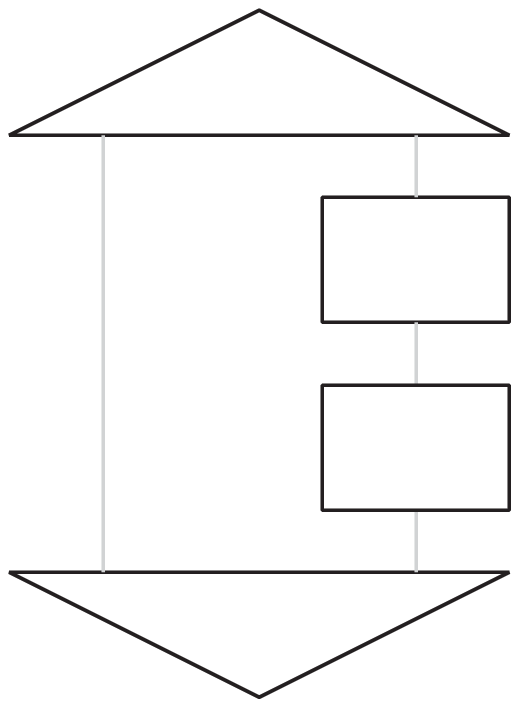,width=70pt}}

\begin{picture}(70,0)
\put(54,72){$f^\dagger$}
\put(54,46){$f$}
\put(33,22){$\eta$}
\put(32,98){$\eta{\!}^\dagger$}
\end{picture}
\end{minipage}}

\vspace{-2mm}
\noindent Bifunctoriality means that we can move these boxes upward and
downward, and naturality provides additional modes of movement
e.g.~scalars admit arbitrary movements --- one could say that they are
not localized in time nor in space but, in Kripke's terms, they \em
provide a weight for a whole world\em. Given that
$f\otimes f^\dagger\!=g\otimes g^\dagger$, that is, in a picture,

\vspace{3mm}\noindent{
\begin{minipage}[b]{1\linewidth}
\centering{\epsfig{figure=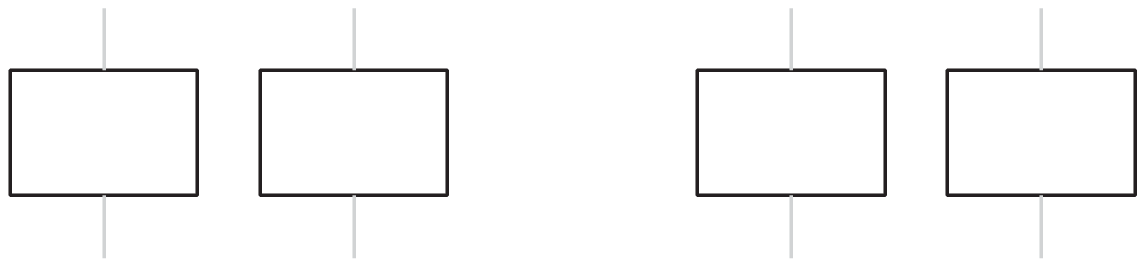,width=157.5pt}}

\begin{picture}(157.5,0)
\put(142,29){$g^\dagger$}
\put(106,29){$g$}
\put(46,29){$f^\dagger$}
\put(10,29){$f$}
\put(71,28){\bf\LARGE=}
\end{picture}
\end{minipage}}

\vspace{-2mm}
\noindent it follows from the picture below that $s\bullet f=t\bullet
g$,

\vspace{3mm}\noindent{
\begin{minipage}[b]{1\linewidth}
\centering{\epsfig{figure=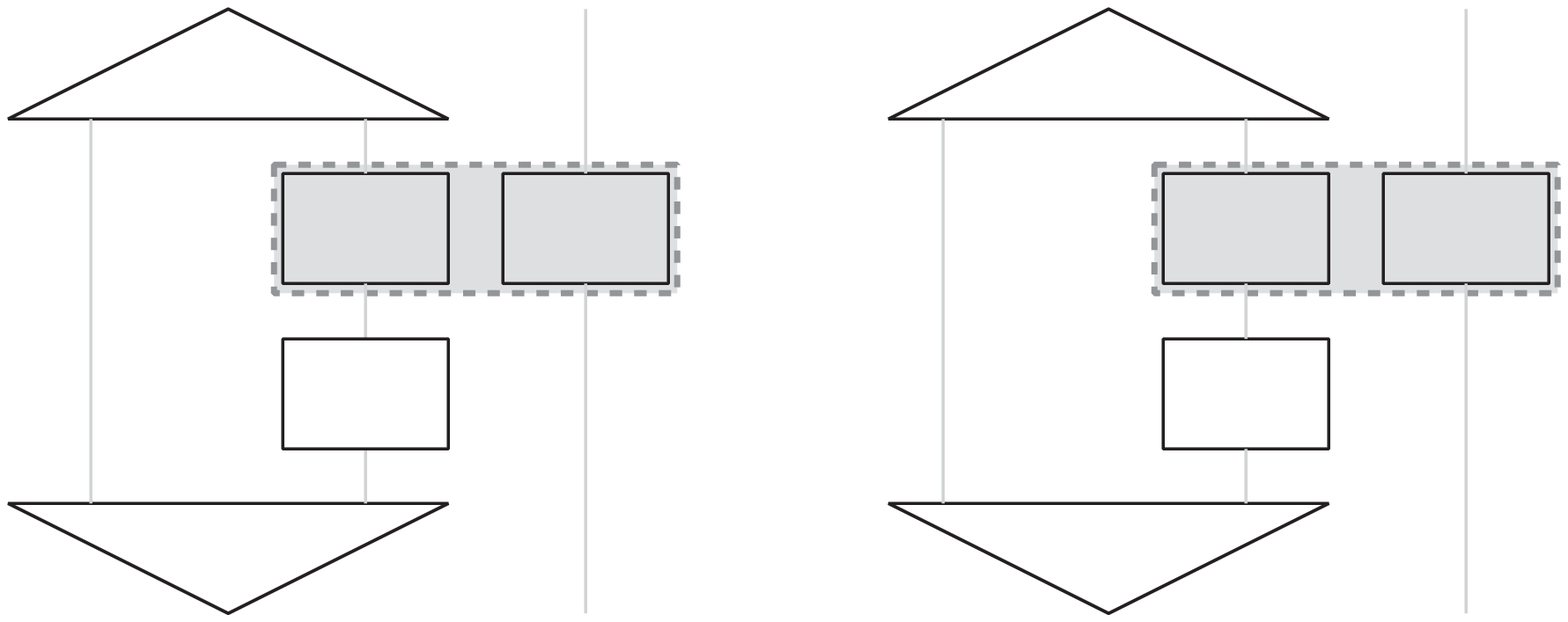,width=245.75pt}}

\begin{picture}(245.75,0)
\put(230,72){$g$}
\put(194,72){$g^\dagger$}
\put(90,72){$f$}
\put(54,72){$f^\dagger$}
\put(194,46){$f$}
\put(54,46){$f$}
\put(33,22){$\eta$}
\put(32,98){$\eta{\!}^\dagger$}
\put(173,22){$\eta$}
\put(172,98){$\eta{\!}^\dagger$}
\put(116,56.5){\bf\LARGE=}
\end{picture}
\end{minipage}}

\vspace{-2mm}
\noindent while the picture below shows that $s\circ s^\dagger=t\circ
t^\dagger$,

\vspace{3mm}\noindent{
\begin{minipage}[b]{1\linewidth}
\centering{\epsfig{figure=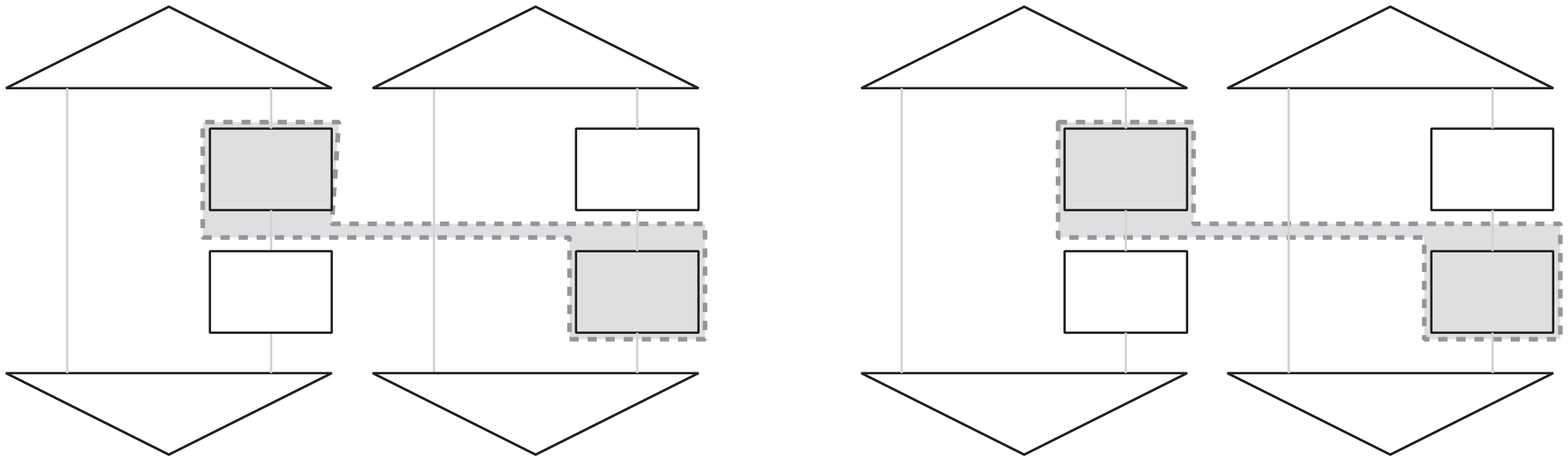,width=333.5pt}}

\begin{picture}(333.5,0)
\put(133,72){$f^\dagger$}
\put(133,46){$f$}
\put(54,72){$f^\dagger$}
\put(54,46){$f$}
\put(33,22){$\eta$}
\put(32,98){$\eta{\!}^\dagger$}
\put(112,22){$\eta$}
\put(111,98){$\eta{\!}^\dagger$}
\put(316,72){$f^\dagger$}
\put(316,46){$g$}
\put(237,72){$g^\dagger$}
\put(237,46){$f$}
\put(216,22){$\eta$}
\put(215,98){$\eta{\!}^\dagger$}
\put(295,22){$\eta$}
\put(294,98){$\eta{\!}^\dagger$}
\put(161,56.5){\bf\LARGE=}
\end{picture}
\end{minipage}}

\vspace{-2mm}

\noindent  This completes the proof of Proposition \ref{Pr:phase2}.

In an SCCC one can also show that
\beq\label{eq:state}
\psi\circ\psi^\dagger=\rho_A^\dagger\circ (\psi\otimes\psi^\dagger)
\circ\lambda_A
\eeq  
where we note that in the case of {\bf FdHilb} the linear
operator
$\psi\circ\psi^\dagger:A\to A$ is the density matrix representing the
pure state
$\psi:\II\to A$ i.e.~the state usually represented by the vector
$\psi(1)\in{\cal H}$. In other words (see
\cite{AC1}), since $\psi$ and $\psi^\dagger$ are respectively  to be conceived
as a ket $|\psi\rangle$ and a bra $\langle\psi|$, their
composite $\psi\circ\psi^\dagger$ corresponds to the ket-bra
$|\psi\rangle\langle\psi|$.  Consider now von Neumann's formalism in
${\bf FdHilb}$. When passing from vectors $\psi$ to density  matrices
$\psi\circ\psi^\dagger$ we cancel out global phases.   The
global amplitudes are squared and hence provide true
probability weights.  This
\em trick
\em however does not extend to morphisms.  Indeed, for $U:A\to B$
unitary we have
$U\circ U^\dagger=1_B$ so we lose all its content.  But
eq.(\ref{eq:state}) tells us that for states we obtain the same
effect (that is eliminating global phases) by passing to
$\psi\otimes\psi^\dagger$ instead of $\psi\circ\psi^\dagger$, and 
this method does extend in abstract generality.  Propositions
\ref{Pr:phase1} and
\ref{Pr:phase2}  then tell us  that the desired effect also extends in
abstract generality for arbitrary morphisms.

Assignments (\ref{eq:Hsscalers}) show that to any morphism, and also
to any pair of morphisms we can attribute a special scalar.  Recall
that  the Hilbert-Schmidt norm of a bounded linear map
$f:{\cal H}_1\to{\cal H}_2$, if it exists, is $\sqrt{\sum_i\langle
f(e_i)\mid f(e_i)\rangle}$ \cite{Dunford}. Such a map which admits a
Hilbert-Schmidt norm is an Hilbert-Schmidt map. When ${\cal H}_1={\cal
H}_2={\cal H}$ all Hilbert-Schmidt maps
${\cal S}({\cal H})$ constitute a Banach algebra with
$\sum_i\langle f(e_i)\mid g(e_i)\rangle$ as an inner-product
\cite{Dunford}.  Hence
${\cal S}({\cal H})$ is itself a Hilbert space.  We still have such a
Hilbert space structure if ${\cal H}_1\not={\cal H}_2$ (we only lose
the compositional structure).

\begin{definition}   For each morphism $f$ in an SCCC {\bf C} we
define its \em squared Hilbert-Schmidt norm
\em as
$||f||:=(\uu f\uuu)^\dagger\circ\uu f\uuu\,\in{\bf C}(\II,\II)$.
\end{definition}

In ${\bf FdHilb}$ we have $||f||(1)=\sum_i \langle f(e_i)\mid
f(e_i)\rangle_{\tilde{\cal H}}$ for
$f:{\cal H}\to\tilde{\cal H}$. For Hilbert spaces ${\cal H}_1$ and
${\cal H}_2\,$, and
${\cal HS}({\cal H}_1,{\cal H}_2)$ the Hilbert space of all
Hilbert-Schmidt maps $f:{\cal H}_1\to{\cal H}_2$, we have
${\cal HS}({\cal H}_1,{\cal H}_2)\,\simeq\,{\cal H}_1\otimes{\cal
H}_2$, so it should not be a surprise that exactly this norm naturally
arises in our setting.

\begin{definition}   
For morphisms $f,g:A\to B$ in an SCCC {\bf C}
we define the \em Hilbert-Schmidt inner-product \em as
$\langle f\mid g\rangle:=(\uu f\uuu)^\dagger\circ\uu g\uuu\,
\in{\bf C}(\II,\II)$.
\end{definition}

Recall from
\cite{AC1,AC1.5} that the \em inner-product of states \em
$\psi,\phi:\II\to A$ in an SCCC is given by
$\psi^\dagger\circ\phi\,\in{\bf C}(\II,\II)$. The Hilbert-Schmidt
inner-product provides a genuine generalization of this inner-product
for states.

\begin{proposition}\label{prop:IPHScorrespondense} For morphisms
$\psi,\phi:\II\to A$ in an SCCC we have
\[ (\uu \psi\uuu)^\dagger\circ\uu\phi\uuu=\psi^\dagger\circ\phi\,.
\]
\end{proposition}

A nice application of Proposition \ref{prop:IPHScorrespondense} is the
derivation of the version of the Born rule which uses the trace and
density matrices. Recall that a projector in the
spectral decomposition attributed to a measurement  decomposes as
$\PP=\pi^\dagger\circ\pi$ and that 
\[
{\rm Prob}(\psi,\PP):=
\psi^\dagger\circ\PP\circ\psi=
(\pi\circ\psi)^\dagger\circ(\pi\circ\psi)
\]
is the corresponding \em abstract probability \em of $\PP$ for measuring
a system in state $\psi:\II\to A$. In the density matrix version of
quantum mechanics \cite{vN} the probability rule is
${\rm Prob}(\rho,\PP):={\rm Tr}(\PP\circ\rho)$ where
$\rho=\psi\circ\psi^\dagger$ is the density matrix corresponding to the
state $\psi$ and
${\rm Tr}$ assigns  to a matrix $(f_{ij})_{ij}$ the trace $\sum_i
f_{ii}$.  Now recall from
\cite{AC1.5} that any strongly compact closed category admits a
categorical
\em partial trace \em in the sense of \cite{JSV} --- this follows
straightforwardly from the corresponding result for compact closed
categories \cite{KellyLaplaza} --- for which the corresponding \em
{\rm(}full{\rm)} trace
\em of $f:A\to A$ is
\[ {\rm Tr}(f):=\eta^\dagger_A\circ (1_{A^*}\otimes
f)\circ\eta_A\,\in{\bf C}(\II,\II)\,.
\] Hence
$\langle f\mid g\rangle={\rm Tr}(f^\dagger\circ g)$.  In {\bf FdHilb}
this categorical trace coincides with the linear algebraic one.
Passing to the Hilbert-Schmidt inner-product through Proposition
\ref{prop:IPHScorrespondense}, applying eq.(\ref{eq:names}),
bifunctoriality and  again eq.(\ref{eq:names}),
\beqa {\rm Prob}(\psi,\PP)
&=&\psi^\dagger\circ(\PP\circ\psi)\\
&=&\eta^\dagger_\II\circ
(1_{\II^*}\otimes\psi^\dagger)\circ(1_{\II^*}\otimes
(\PP\circ\psi))\circ\eta_\II\\
&=&\eta^\dagger_A\circ ((\psi^\dagger)^*\otimes
1_{A})\circ(1_{\II^*}\otimes
(\PP\circ\psi))\circ\eta_\II\\
&=&\eta^\dagger_A\circ (1_{A^*}\otimes
(\PP\circ\psi))\circ((\psi^\dagger)^*\otimes
1_\II)\circ\eta_\II\\ &=&\eta^\dagger_A\circ (1_{A^*}\otimes
(\PP\circ\psi))\circ(1_{A^*}\otimes\psi^\dagger)
\circ\eta_A\,=\,{\rm Tr}(\PP\circ\rho)\,.
\eeqa  But in a picture all this boils down to merely moving
$\psi^\dagger$ around a loop:

\vspace{3mm}\noindent{
\begin{minipage}[b]{1\linewidth}
\centering{\epsfig{figure=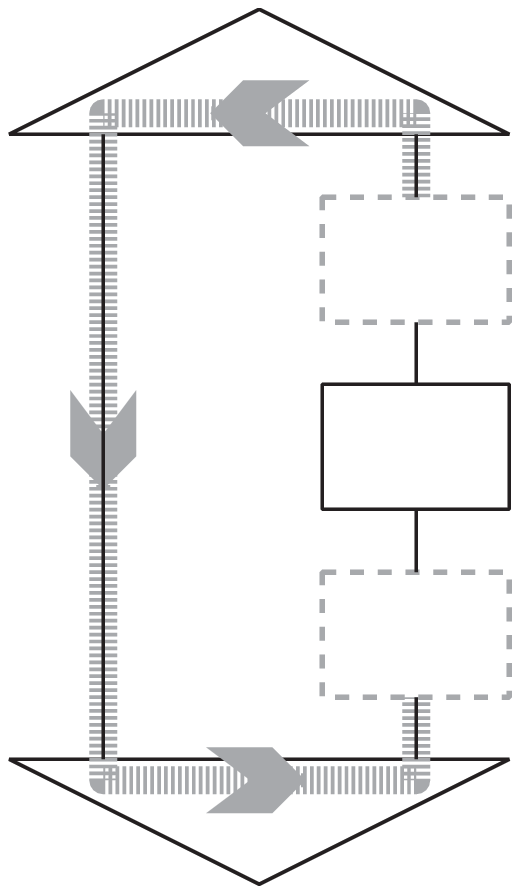,width=105pt}}

\begin{picture}(105,0)
\put(81,142.2){$\psi^\dagger$}
\put(67.5,102.7){$\ \PP\circ\psi$}
\put(81,63.2){$\psi^\dagger$}
\put(49,21){$\eta$}
\put(48,185){$\eta{\!}^\dagger$}
\end{picture}
\end{minipage}}

\vspace{-2mm}

\section{The preparation-state agreement
axiom}\label{sec:elliminateph}

Following Propositions \ref{Pr:phase1} and
\ref{Pr:phase2} the following construction aims at eliminating global
phases i.e.~it tries to turns a category with \em vector space
flavored \em objects into one with
\em projective space `with weights' flavored \em objects.

\begin{definition}\label{def:QL} For each SCCC ${\bf C}$ we define a
category
${\it WProj}({\bf C})$.
\bit
\item The objects of ${\it WProj}({\bf C})$ are those of ${\bf C}$.
\item The {\rm Hom-}sets of ${\it WProj}({\bf C})$ are
\beq\label{eq:WPROJconst} {\it WProj}({\bf C})(A,B):=\{f\otimes
f^\dagger\mid f\in{\bf C}(A,B)\}
\eeq with $1_A\otimes 1_A\in {\it WProj}({\bf C})(A,A)$ being the
identities.
\item Composition in ${\it WProj}({\bf C})$, for $A\rTo^f B\rTo^g C$
in
${\bf C}$, is given by
\[ (f\otimes f^\dagger)\,\bar{\circ}\,(g\otimes g^\dagger):=  (f\circ
g)\otimes(f\circ g)^\dagger\!.
\]
\eit
\end{definition}

\begin{proposition}\label{pr:QL} Let ${\bf C}$ be an SCCC.  Then ${\it
WProj}({\bf C})$ is also an SCCC.
\end{proposition}

\begin{proposition}\label{pr:altcharac} If $f$ and $g$ are morphisms
in an SCCC then
\[ f\otimes f^\dagger\!=g\otimes g^\dagger
\ \ \Longleftrightarrow\ \  f\otimes f_*=g\otimes g_*
\ \ \Longleftrightarrow\ \  {\rm P}_f={\rm P}_g\,.
\]
\end{proposition}

While one easily verifies that $A\mapsto A$, $f\otimes
f^\dagger\mapsto f$ yields
\[ ({\it WProj}\circ{\it WProj})({\bf FdHilb})\simeq{\it WProj}({\bf
FdHilb})\,,
\] idempotence of ${\it WProj}$ fails to be true for arbitrary SCCC.
Hence we are mainly interested in \em invariance \em (up to
isomorphism) under
${\it WProj}$.

\begin{theorem}\label{prop:idempwprof}
${\it WProj}({\bf C})\simeq{\bf C}$ {\rm(}canonically{\rm)} for an SCCC
${\bf C}$ if and only if
\beq\label{eq:idempQL} f\otimes f^\dagger=g\otimes g^\dagger\
\Longrightarrow\ f=g\,.
\eeq
\end{theorem}

Condition (\ref{eq:idempQL}) expresses that an SCCC is a \em fixed point
\em of the {\it WProj\,}-construction --- as our main example ${\it
WProj}({\bf FdHilb})$ is one --- and hence it guarantees absence of
redundant global phases. Roughly speaking one can think  of these fixed
points  as being the result of consecutively applying ${\it WProj}$
until all global redundancies are erased. But condition
(\ref{eq:idempQL}) also admits a lucid interpretation in its own right
which is moreover a truly compelling physical motivation to adopt
condition (\ref{eq:idempQL}) as an
\em axiom \em for any categorical model of abstract quantum mechanics.

\begin{corollary}
${\it WProj}({\bf C})\simeq{\bf C}$ {\rm(}canonically{\rm)} for an SCCC
${\bf C}$ if and only if
\beq\label{eq:idempQL3} {\rm P}_f={\rm P}_g\ \Longrightarrow\ \uu
f\uuu=\uu g\uuu.
\eeq
\end{corollary}

Condition (\ref{eq:idempQL3}) states that if  two preparations ${\rm
P}_f$ and ${\rm P}_g$ of bipartite states coincide then we have (of
course) that the bipartite states $\uu f\uuu$ and $\uu g\uuu$ which
they produce coincide.  And without loss of generality this fact
extends to arbitrary states --- recall here form \cite{AC1} that
$\psi\circ\psi^\dagger:A\to A$ is the projector which prepares the
state
$\psi:\II\to A$.

\begin{corollary}\label{prop:Fromftopsi}
${\it WProj}({\bf C})\simeq{\bf C}$ {\rm(}canonically{\rm)} for an SCCC
${\bf C}$ if and only if
\beq\label{eq:idempQL4}
\forall\ \psi,\phi:\II\to A\,,\ \ \
\psi\circ\psi^\dagger=\phi\circ\phi^\dagger\ \Longrightarrow\
\psi=\phi\,.
\eeq
\end{corollary}

\begin{definition}\label{def:preparation-state} An SCCC satisfies
the
\em preparation-state agreement axiom \em iff the equivalent
conditions  (\ref{eq:idempQL}), (\ref{eq:idempQL3}) and
(\ref{eq:idempQL4}) are satisfied.
\end{definition}

\section{Ortho-structure}\label{sec:orthostructure}

Besides being an SCCC {\bf FdHilb} also has \em biproducts \em i.e.~it
is
\em semi-additive\em.  For an SCCC with biproducts the endomorphism
monoid of the tensor unit is always an \em involutive abelian
semiring\em, and the full subcategory of objects of type
$\II\oplus\ldots\oplus\II$ is isomorphic to the category of matrices
in that involutive abelian semiring, and conversely, each \em matrix
calculus \em over an involutive abelian semiring provides an example
of an SCCC with biproducts \cite{AC1}.

\begin{theorem}  There exist no SCCC with biproducts which both
satisfies the preparation-state agreement axiom and for which the
endomorphism monoid of the tensor unit is a ring with non-trivial
negatives {\rm(}i.e.~$-1=1${\rm)}.
\end{theorem}

So if a category with as morphisms matrices over a commutative
involutive  semiring $R$ satisfies the preparation-state agreement
axiom then $R$ cannot have non-trivial negatives, with the fatal
consequence that \em interference phenomena \em relying on
cancellation of negatives cannot be modeled.  Note that our key
example ${\it WProj}({\bf FdHilb})$ is not isomorphic to the matrix
calculus over its scalar monoid
$\mathbb{R}^+\!$.  Next set
$[f]:=\bigl\{g\bigm| f\otimes f^\dagger=g\otimes g^\dagger\bigr\}$.

\begin{theorem}\label{eq:monoidalfailure} 
The product structure and the
symmetric monoidal $-\oplus-$ structure of ${\bf FdHilb}$ do \underline{not}
carry over to ${\it WProj}({\bf FdHilb})$. In particular, in an SCCC
with biproducts $f'\in[f]$ and
$g'\in[g]$ do \underline{not} imply $f'\oplus g'\in [f\oplus g]$ and
hence the operation $[f]\,\bar{\oplus}\,[g]:=[f\oplus g]\,$ is
\underline{ill}-defined.
\end{theorem}
\bpf While $1_\mathbb{C}\in[1_\mathbb{C}]$ and
$(e^{i\theta}\!\circ-)\in[1_\mathbb{C}]$ we have
$\langle 1_\mathbb{C},(e^{i\theta}\!\circ-)\rangle\not\in [\langle
1_\mathbb{C}, 1_\mathbb{C}\rangle]$ and
$1_\mathbb{C}\oplus(e^{i\theta}\!\circ-)\not\in[1_\mathbb{C}\oplus
1_\mathbb{C}]\,$.
\endproof\newline

There is a physical argument why we do not want a product structure. A
pairing  operation
$\langle-,-\rangle:{\bf C}(\II,\II)\times{\bf C}(\II,\II)\to{\bf
C}(\II,\II\oplus\II)$ would allow to deduce the initial state
$\psi:\II\to\II\oplus\II$ of a \em qubit
\em from the probabilities
\[ {\rm Prob}\bigl(\psi\,,\,p_{\underline{\II},\II}^\dagger\!\circ
p_{\underline{\II},\II}\bigr)\,\in {\bf C}^+(\II,\II)
\qquad{\rm and}\qquad {\rm
Prob}\bigl(\psi\,,\,p_{\II,\underline{\II}}^\dagger\!\circ
p_{\II,\underline{\II}}\bigr)\,\in {\bf C}^+(\II,\II)
\] of it being subjected to the dichotomic measurement with projectors
\[ p_{\underline{\II},\II}^\dagger\!\circ
p_{\underline{\II},\II}:\II\oplus\II\to\II\oplus\II
\qquad\quad{\rm and}\qquad\quad
p_{\II,\underline{\II}}^\dagger\!\circ
p_{\II,\underline{\II}}:\II\oplus\II\to\II\oplus\II
\] since
$\psi=\bigl\langle {\rm
Prob}\bigl(\psi\,,\,p_{\underline{\II},\II}^\dagger\!\circ
p_{\underline{\II},\II}\bigr)\,{\bf,}\ {\rm
Prob}\bigl(\psi\,,\,p_{\II,\underline{\II}}^\dagger\!\circ
p_{\II,\underline{\II}}\bigr)\bigr\rangle$.  But this contradicts the
\em empirical evidence \em that a qubit state comprises
\em relative phase data \em --- which is responsible for interference
phenomena --- which gets erased by a measurement.  So the intrinsic
\em informatic irreversibility\,\em\footnote{Not to be confused with
the irreversibility of a projector as a linear map.} of quantum
measurements clashes with the very nature of the concept of a
categorical  product.

Part of the symmetric monoidal structure can actually be retained. The
problem exposed in  Theorem \ref{eq:monoidalfailure} can be overcome
if both
$[f]$ and
$[g]$ contain a particular distinguished morphism, say respectively
$f$ and
$g$ themselves.  Then we can define their monoidal sum by setting
$[f]\,\bar{\oplus}\,[g]:=[f\oplus g]$. There are important equivalence
classes which have such a distinguished element:
\bit
%\item \em Zero maps \em $0_{A,B}:A\to B$ in $[0_{A,B}]=\{0_{A,B}\}$.
\item \em Identities \em $1_A:A\to B$ in $[1_A]$ as a  part of the
categorical structure.
\item \em Natural isos \em $\lambda_A, \rho_A,
\sigma_{A,B},\alpha_{A,B,C}, u_\II$ as part of the SCCC structure.
\item \em Positive scalars \em $s\circ s^\dagger:\II\to\II$ whenever
they are unique in $[s\circ s^\dagger]$.
\eit Such distinguished morphisms are the only ones for which we need monoidal
sums, so we are going to let them play a distinguished role within the
`minimally required' additive structure which we will introduce.  Indeed, much
of what seems to be additive at first sight turns out to be
multiplicative e.g.~while the usual Hilbert-Schmidt norm involves an
explicit
\em summation \em of inner-products parametrized over a basis,
abstractly it only involves units and adjoints which are both part of
the multiplicative SCCC-structure.

\begin{definition}\label{def:orthostruc}  An \em ortho-SCCC \em is an SCCC
$({\bf C},\otimes, \II, \lambda,\rho,\sigma, \alpha,(-)^*,
(-)^\dagger,\eta)$ which comes with an
\em ortho-structure \em i.e.~a second monoidal structure $(-\oplus-)$ which is total on objects but can be only partial on morphisms, more specifically, the symmetric monoidal category
$({\bf C}^\sharp,\oplus, 0,l,r,s,a)$ is a subcategory of ${\bf C}$ of which the objects coincide with those of ${\bf C}$, and,
which is such that $(-\otimes-):{\bf C}^\sharp\times{\bf
C}^\sharp\to{\bf C}^\sharp$ is a strong symmetric monoidal bifunctor of
which the witnessing natural isomorphisms are unitary and with
$\lambda,\rho,\alpha,\sigma,u_\II$ symmetric monoidal natural in all
variables, and which is also such that the partial bifunctor
$(-\oplus-)$ commutes with
$(-)^*$ and
$(-)^\dagger$ i.e.
$0^*=0$, $(A\oplus B)^*=A^*\oplus B^*$ for all objects and
$(f\oplus g)^\dagger=f^\dagger\oplus g^\dagger$ for all ${\bf
C}^\sharp$-morphisms.
\end{definition}

The strong symmetric monoidal
bifunctor provides (by definition) \em distributivity \em natural isomorphisms
\[
\dist_{0,l}:A\otimes 0\simeq 0\qquad\qquad \dist_l:  A\otimes(B\oplus
C)\simeq(A\otimes B)\oplus(A\otimes C)\,\,
\]
\[
\dist_{0,r}:0\otimes A\simeq 0\qquad\qquad \dist_r: (B\oplus C)\otimes
A\simeq(B\otimes A)\oplus(C\otimes A)\,.
\] 
By asking that $(-\otimes-)$ is a strongly symmetric monoidal bifunctor with
$\lambda,\rho,\sigma,\alpha,u_\II$ all monoidal natural isomorphisms we make sure
that these distributivity isomorphisms behave well with respect to the natural
isomorphisms of the symmetric monoidal structure on ${\bf C}^\sharp$.  

\begin{proposition}\label{Oenrich}
For each pair of objects $A,B$ in an ortho-SCCC there exists a distinguished morphism $0_{A,B}:A\to B$ which, for all $f:B\to C$ satisfies
\[
f\circ 0_{A,B}=0_{B,C}\circ f=0_{A,C}
\]
and which is explicitly defined in 
\begin{diagram}
A&\rTo^{\hspace{-2.5cm}\simeq\hspace{-2cm}}&{\rm I}\otimes
A&\rTo^{\eta_0\otimes 1_A\!\!}&(0^*\!\otimes 0)\otimes
A&\rTo^{\simeq\!\!\!\!\!\!\!\!\!\!}&0\\
\dTo^{0_{A,B}}&&&&&&\dTo_{1_0}\\
B&\lTo_{\hspace{-2.5cm}\simeq\hspace{-2cm}}&{\rm I}\otimes
B&\lTo_{\eta_0^\dagger\otimes 1_B\!\!}&(0^*\!\otimes 0)\otimes
B&\lTo_{\simeq\!\!\!\!\!\!\!\!\!\!}&0
\end{diagram}
\end{proposition}
 
Hence, without assuming the universal property of a zero object
we do obtain a special family of morphisms which behave similarly. We will set $0_A:=0_{A,0}$, hence $0_{A,B}:=0^\dagger_B\circ 0_A:A\to B$.  We define \em pseudo-projections
\em and the \em  pseudo-injections \em respectively as
\[ 
p_{\underline{A},B}:=r_A^\dagger\!\circ(1_A\oplus 0_B):A\oplus B\to A
\]
\[
p_{A,\underline{B}}:=l_A^\dagger\!\circ(0_A\oplus 1_B):A\oplus
B\to B
\]
\[ q_{\underline{A},B}:=(1_A\oplus 0_B^\dagger)\circ r_A:A\to A\oplus B
\]
\[
q_{A,\underline{B}}:=(0_A^\dagger\oplus 1_B)\circ l_A:B\to
A\oplus B
\] and the \em pseudo-components \em of a morphism $f:\bigoplus_i
A_i\to
\bigoplus_j B_j$ are
\[ 
f_{ij}:=p_j\circ f\circ q_i:A_i\to B_j\,.
\] 
Of course in general these do \underline{not} admit any kind of matrix calculus.

\begin{proposition} In an ortho-SCCC we have
\[
\ p_{\underline{A},B}\circ q_{\underline{A},B}=1_A
\qquad\qquad\qquad\qquad\quad
\!\!\!\!p_{\underline{A},B}\circ q_{A,\underline{B}}=0_{A,B}\ \
\]
\[
\
\!\!\!\!\!\!q_{\underline{A},B}^\dagger=p_{\underline{A},B}=
p_{B,\underline{A}}\circ s_{A,B}\quad\quad
\qquad\quad p_{\underline{B},D}\circ(f\oplus g)=f\circ
p_{\underline{A},C}
\]
\[
\ 1_A\oplus p_{\underline{B},C}=p_{\underline{A\oplus B},C}\circ
a_{A,B,C}
\qquad\quad p_{\underline{A},B}\circ p_{\underline{A\oplus B},C}=
p_{\underline{A},B\oplus C}\circ a^\dagger_{A,B,C}\,.\!\!
\]
\end{proposition}

\begin{proposition} 
The components
$\pi_i:=p_i\circ U:A\to A_i$ of a unitary morphism $U:A\to\bigoplus_i
A_i$ are `conormalized' i.e.~$\pi_i\circ\pi_i^\dagger=1_{A_i}$ and
`coorthogonal' for $i\not=j$
i.e.~$\pi_j\circ\pi_i^\dagger=0_{A_i,A_j}$. Analogously,  the
components $\psi_i:=U\circ q_i:A_i\to A$ of unitary morphism
$U:\bigoplus_i A_i\to A$ are `normalized'
i.e.~$\psi^\dagger_i\circ\psi=1_{A_i}$ and `orthogonal' for
$i\not=j$ i.e.~$\psi_j^\dagger\circ\psi_i=0_{A_i,A_j}$. Unitary maps
preserve normality, conormality, orthogonality and coorthogonality.
\end{proposition}

Note that the partial monoidal sum on morphisms did not come with an
operational significance since its only aim was to provide
pseudo-projections and pseudo-injections with appropriate properties.
But they do much more than this, they also provide sums.

\begin{theorem}\label{thm:defsum}
An ortho-SCCC is (partially) enriched in commutative monoids i.e.~admits a notion of sum of morphisms, where the sum of  $f,g:A\to B$ for which $f\oplus g$ exists is given by
\begin{diagram}
B&\lTo^{\hspace{-2.5cm}\simeq\hspace{-2cm}}&{\rm I}\otimes
B&\lTo^{\eta_2^\dagger\otimes 1_B\!\!}&(2^*\!\otimes 2)\otimes
B&\lTo^{\simeq}&2^*\!\otimes(B\oplus B)\\
\uTo_{f+g}&&&&&&\uTo~{1_{2^*}\!\otimes(f\oplus g)}\\
A&\rTo_{\hspace{-2.5cm}\simeq\hspace{-2cm}}&{\rm I}\otimes
A&\rTo_{\eta_2\otimes 1_A\!\!}&(2^*\!\otimes 2)\otimes
A&\rTo_{\simeq}&2^*\!\otimes(A\oplus A)
\end{diagram}
and with the additive units as in Proposition \ref{Oenrich}.
\end{theorem}

\smallskip\noindent
We can put this in a slogan:

\medskip\noindent
\begin{center}
\fbox{SCCC + $\oplus$ + distributivity $\ \ \Longrightarrow\ \ $ {\bf CMon}-enrichment}
\end{center}

\section{Categorical semantics for protocols}\label{sec:categoricalsem}

An ortho-SCCC provides enough structure for the description and
correctness proofs of the protocols considered in \cite{AC1}. Two
approaches are possible.

\paragraph{5.1. Distinct types for superposition and weighted
branching.}  When starting from an ortho-SCCC it suffices to add
classical branching \em freely \em as a product structure i.e.~\em sum
types \em
$(A_1,\ldots,A_n)$, \em pairing \em $\langle
f_1,\ldots,f_n\rangle:C\to(A_1,\ldots,A_n)$ and \em projections \em
$\tilde{p}_i:(A_1,\ldots,A_n)\to A_i$. This \em branching structure \em
enables classical statistics and measurement outcome dependent
manipulation of data i.e.~\em classical information flow\em, while the
ortho-structure provides the interface between the quantum state space
and the classical world. We adapt some examples from \cite{AC1} to the
context of an ortho-SCCC with freely added products. Each unitary
morphism
${U:A\to \bigoplus_{i=1}^{i=n}A_i}$ defines a \em non-destructive
measurement \em
$\langle \PP_i\rangle_{i=1}^{i=n}:A\to (A)_{i=1}^{i=n}$ where
$\PP_i:=\pi^\dagger_i\circ\pi_i$ with $\pi_i:=p_i\circ U$.  While in
\cite{AC1} classical communication is encoded as distributivity
isomorphisms here we have
\[ {\rm CC}_{A\leftarrow(B,C)}:=\left\langle
1_A\otimes\tilde{p}_1,1_A\otimes\tilde{p}_2\right\rangle:
A\otimes(B,C)\to(A\otimes B,A\otimes C)
\]
which admits no inverse, reflecting the fact that in
absence of the ability to erase information, distributing information
is irreversible. Also, while there is a canonical map
$\langle p_{\underline{\II},\II},p_{\II,\underline{\II}}\rangle:
\II\oplus\II\to(\II\,,\II)$, namely the destructive measurement
associated to the unitary morphism
$1_{\II\oplus\II}$, this map has no inverse, and hence there exists no
isomorphism between a \em qubit \em $Q\simeq\II\oplus\II$ and a \em
weighted bit
\em $(\II,\II)$. More concretely, we define a \em destructive
teleportation measurement
\em by means of a unitary morphism
$T:Q\otimes Q\to \II\oplus\II\oplus\II\oplus\II$ which is such that
there exist unitary maps
$\beta_1,\beta_2,\beta_3,\beta_4:Q\to Q$ with
$\uu \beta_i\uuu=T^\dagger\circ q_i$. The destructive teleportation
measurement itself is
\[
\langle p_i\circ T\rangle_{i=1}^{i=4}=\langle \uu
\beta_i\uuu^\dagger\rangle_{i=1}^{i=4}:Q\otimes Q\to
(\II,\II,\II,\II)\,.
\]

\begin{theorem} The theorems stated in \em \cite{AC1} \em on
correctness of the example protocols for an SCCC with biproducts carry
over to any ortho-SCCC with freely added products when using the above
definitions.
\end{theorem}

Hence it indeed suffices for the ortho-structure to be limited to
assuring coherent coexistence of the pseudo-projections with the SCCC
structure since   for all the other qualitative uses of the biproduct
structure in
\cite{AC1} we can as well use the freely added product structure which
does not genuinely interact with the SCCC structure. Conclusively, we
decomposed the additives in a \em fundamental structural component\em,
namely the ortho-structure, and, a \em classical branching
structure\em, which can be freely added as a product structure. This
classical branching structure can of course be of a more sophisticated
nature than the one we used here, for example one might want to
capture classical mixing, but the bottom line is that it can be
introduced on top of the ortho-SCCC structure and hence is not an
intrinsic ingredient.

\paragraph{5.2. \cite{AC1}-style semantics.} One keeps a
minimal number of non-isomorphic types  by distinguishing between
explicit and non-explicit sums. For example, when $Q\simeq \II\oplus
\II$ then $Q$ represents a qubit i.e.~the \em superposition \em of
$\II$ and $\II$, while
$\II\oplus \II$ represents a pair of \em probabilistic weights
\em attributed to two \em branches \em of scalar type e.g.~the
respective probabilities of a destructive non-trivial qubit
measurement. Since this semantics is discussed in detail in \cite{AC1}
we only point at the required modification when starting from an
ortho-SCCC rather than from an SCCC with biproducts.  The key
observation is that given a unitary morphism
$U:A\to \bigoplus_{i=1}^{i=n}A_i$ with $q_i:A_i\to A$
pseudo-injections we can define the corresponding non-destructive
measurement as
\[
\left(\bigoplus_{i=1}^{i=n}
U^\dagger\right)\circ\left(\bigoplus_{i=1}^{i=n} q_i\right)\circ
U:A\to\bigoplus_{i=1}^{i=n}A
\]   which in the case of biproducts coincides with
$\langle \pi^\dagger_i\circ\pi_i\rangle_{i=1}^{i=n}:A\to
\bigoplus_{i=1}^{i=n}A$.    Note that $q_i\in{\bf C}^\sharp$ and that
it is also reasonable to assume meaningfullness of
$\bigoplus_{i=1}^{i=n} U^\dagger$ i.e.~$n$ copies of the same
morphism. However, brench dependent operations $\bigoplus_{i=1}^{i=n} f_i$ do require a sufficiently large additive monoidal structure.  
The state of the $j$th branch is obtained by applying 
\[
p_j\circ-:{\bf C}\!\left(A,\bigoplus_{i=1}^{i=n}A_i\right)\to{\bf C}(A,A_j)\,,
\]
and the absence of a pairing operation as in ${\it WProj}({\bf FdHilb})$ prevents the
collapse of the classical information flow onto the superposition structure.

\section{What is a Born-rule?}\label{sec:Born-rule}

Given a model which intends to describe quantum mechanics, including
\em states \em ${\cal S}$, \em measurements \em
${\cal M}$, \em probability weights \em
$\mathbb{W}$, a \em unit \em $1\in\mathbb{W}$ and an addition
$-+-:\mathbb{W}\times\mathbb{W}\to\mathbb{W}$, a
\em Born rule
\em assigns to each tuple consisting of a state $\psi\in{\cal S}$, a
measurement
$M\in{\cal M}$ which applies to $\psi$ and an outcome $i\in {\rm
spec}(M)$ a probability weight $s(\psi,M,i)\in\mathbb{W}$ such that
\[
\sum_{i\in {\rm spec}(M)}\!\!\!\!s(\psi,M,i)=1\,.
\] If our model is both \em compositional \em and if states \em carry a
probability weight \em which can be extracted by means of a \em
valuation
\em
$|-|_\xi:{\cal S}\to\mathbb{W}$ we obtain
\[
\sum_i|M_i\circ\psi|_\xi=|\psi|_\xi
\]
where $M_i\circ -$ stands for
the \em action \em of $M$ on the state $\psi$ whenever the outcome $i$
occurs in that measurement. E.g.~in {\bf FdHilb} we have
$|\psi|_{\mbox{\rm\bf\tiny FdHilb}}:=\langle\psi\mid\psi\rangle$ for
$\psi\in{\cal H}$ and
$M_i:=\PP_i:{\cal H}\to {\cal H}$ for the non-destructive measurement
represented by the self-adjoint operator
$M=\sum_ia_i\cdot\PP_i:{\cal H}\to {\cal H}$, so since
$|\PP_i\circ\psi|_{\mbox{\rm\bf\tiny
FdHilb}}=\langle\PP_i\circ\psi\mid\PP_i\circ\psi\rangle=
\langle\psi\mid\PP_i\circ\psi\rangle$ by self-adjointness of $\PP_i$
and since $\sum_i\PP_i=1_{\cal H}$ the usual Born rule slightly
generalized to the case that
$|\psi|_{\mbox{\rm\bf\tiny FdHilb}}\not=1$ arises i.e.
\[
\sum_i|\PP_i\circ\psi|_{\mbox{\rm\bf\tiny FdHilb}}=
|\psi|_{\mbox{\rm\bf\tiny FdHilb}}\,.  
\]
Using
$\PP_i=\pi^\dagger_i\circ\pi_i$ with
$\pi_i:{\cal H}\to{\cal G}_i$ and hence
${\cal H}\simeq\bigoplus_i{\cal G}_i$ we have
$|\PP_i\circ\psi|_{\mbox{\rm\bf\tiny
FdHilb}}=|\pi_i\circ\psi|_{\mbox{\rm\bf\tiny FdHilb}}$, and for
$U:{\cal H}\to \bigoplus_i{\cal G}_i\,$ the unique unitary map
satisfying $p_i\circ U=\pi_i$, where
$p_j:\bigoplus_i{\cal G}_i\to{\cal G}_j$ are the canonical
projections, when setting
$\phi:=U\circ\psi:\mathbb{C}\to \bigoplus_i{\cal G}_i$  we obtain
$|\psi|_{\mbox{\rm\bf\tiny FdHilb}}=|\phi|_{\mbox{\rm\bf\tiny
FdHilb}}$. When also introducing the
\em components \em of $\phi$ as
$\phi_i:=p_i\circ\phi=p_i\circ U\circ\psi=\pi_i\circ\psi:\mathbb{C}\to
{\cal G}_i$ all the above results in
\[
\sum_i|\phi_i|_{\mbox{\rm\bf\tiny FdHilb}}= |\phi|_{\mbox{\rm\bf\tiny
FdHilb}}\,.
\] 
When replacing the \em squared vector norm \em
$|-|_{\mbox{\rm\bf\tiny FdHilb}}=\langle-\mid-\rangle:({\cal
H}\to\mathbb{C})\to\mathbb{C}$ which only applies to morphisms of the
type
$\mathbb{C}\to{\cal H}$ by the
\em squared Hilbert-Schmidt norm \em
$|-|_{\mbox{\rm\bf\tiny
FdHilb}}=\sum_i\bigl\langle-(e_i)\bigm|-(e_i)\bigr\rangle:({\cal
H}_1\to{\cal H}_2)\to\mathbb{C}$ we obtain   by an analogous
calculation that
\beq\label{eq:BornpostHS}
\sum_i|f_i|_{\mbox{\rm\bf\tiny  FdHilb}}= |f|_{\mbox{\rm\bf\tiny
FdHilb}}\
\eeq where now
$f:{\cal H}\to\bigoplus_i{\cal G}_i$ and $f_i:=p_i\circ f:{\cal
H}\to{\cal G}_i$ with ${\cal H}$ arbitrary.  For obvious reasons
eq.(\ref{eq:BornpostHS}) is our favorite incarnation of the orthodox
Born rule.

Hence expressing a Born rule requires a \em scalar sum
\em
${-+-}$ and a scalar-valued
\em valuation \em on morphisms $|-|_\xi$, and when interpreting
scalars as probabilistic weights these respectively stand for
\em adding probabilities \em and \em extracting \em the \em
probabilistic weight \em from the morphisms representing physical
processes. The Born rule itself should then express that `taking
components of morphisms', that is, physically speaking, `branching due
to measurements', reflects through the valuation at the level of the
scalars in terms of a decomposition over the scalar sum,
diagrammatically,
\begin{diagram}
\,{\bf C}(A,B_1\oplus\ldots\oplus B_n)&\rTo^{\ \ \left\langle
p_1\circ-\,,\
\ldots\,,\,p_n\circ-\right\rangle_{\bf Set}\
\ }&{\bf C}(A,B_1)\times\ldots\times{\bf C}(A,B_n)\\
\dTo_{|-|_\xi}&&\dTo^{\,|-|_\xi\times\
\ldots\,\times|-|_\xi}\\ {\bf C}(\II,\II)&\lTo_{-+\,\ldots\,+-}&{\bf
C}(\II,\II)\times\ldots\times{\bf C}(\II,\II)\,.
\end{diagram}  
Physically this means that the total probability weight
is preserved when considering all branches i.e.~a \em conservation
law\em, which in particular implies that
\em relative phases lost in measurements carry no probabilistic
weight\em.

For $f:A\to B$ and $h:A\to A$ we (re-)set
\[
\ ||f||:=\eta^\dagger_A\circ(1_{A^*}\otimes (f^\dagger\circ
f))\circ\eta_A
\quad\ {\rm and} \quad\ {\rm Tr}(h):=\eta^\dagger_A\circ(1_{A^*}\otimes
h)\circ\eta_A\,.
\] The \em trace \em ${\rm Tr}$ only applies to
\em endomorphic types \em and
$||f||={\rm Tr}(f^\dagger\circ f)$. In an SCCC with biproducts the
trace is
\em linear \em i.e.~${\rm Tr}(h+h')={\rm Tr}(h)+{\rm Tr}(h')$.

\begin{proposition}\label{pr:biprodgenBOern} Each SCCC with biproducts
admits a Born rule, namely,
$||f_1||+\ldots +||f_n||=||f||$ for any morphism $f:A\to
B_1\oplus\ldots\oplus B_n$.
\end{proposition}

We set $||{\bf C}||$ for the range of $||-||$ and $|f|:=\sqrt{||f||}$
if
$||f||$ has a unique square-root. We take a scalar to be \em positive
\em iff it decomposes as $x\circ x^\dagger$ and has at most one
square-root in
\[
{\bf C}^+(\II,\II):=\{s^\dagger\circ s \mid s\in {\bf C}(\II,\II)\}\,
\]
and an SCCC {\bf C} to be \em positive valued
\em iff all the scalars in $||{\bf C}||$ are positive.

\begin{proposition}\label{pr:BornfromWPBP} If ${\bf C}$ is a positive
valued SCCC with biproducts then
\[ |f_1|_{{\it WProj}({\bf C})}+\ldots+|f_n|_{{\it WProj}({\bf
C})}=|f|_{{\it WProj}({\bf C})}
\] for any $f\in{\it WProj}({\bf C})(A,B_1\oplus\ldots\oplus B_n)$ so
${\it  WProj}({\bf C})$ admits a Born rule.
\end{proposition}

When does a non-semi-additive ortho-SCCC admit a Born rule?  Does it
matter that the valuation involves $||-||$ i.e.~relies on the
multiplicative structure? Is $||-||$
(cf.~Prop.~\ref{pr:biprodgenBOern}) or $|-|$
(cf.~Prop.~\ref{pr:BornfromWPBP}) more canonical than the other? The
following lemma shows that if ${{\bf C}(\II,\II)\subseteq{\bf
C}^\sharp(\II,\II)}\,$, then for any valuation which is a rational
power
$\nu$ of
$||-||$ there is a single structural axiom which stands for existence
of a Born-rule, and which only relies on the SCCC-structure and on
$-\oplus-$ (hence not explicitly on $\nu$ nor on $-+-$).

\begin{lemma}\label{superlemma} Let ${\bf C}$ be an ortho-SCCC and let
the maps
\[ |-|_\xi:\bigcup_{A,B}{\bf C}(A,B)\to{\bf C}(\II,\II)
\quad\ {\rm and}\quad\ -+-:{\bf C}(\II,\II)\times{\bf
C}(\II,\II)\to{\bf C}(\II,\II)\,,
\] be such that for all morphisms $f:A\to B_1\oplus B_2$ we have
\beq\label{eq:OBvaluation} |f|_\xi=|f_1|_\xi+|f_2|_\xi\,.
\eeq
\ben
\item[{\bf i.}] Scalars $s_1, s_2, s_3$ satisfy an associative rule
$(s_1+s_2)+s_3=s_1+(s_2+s_3)$ provided there exists a morphism $f:A\to
B_1\oplus B_2\oplus B_3$ such that $s_i=|f_i|_\xi$. Hence
$f:A\to B_1\oplus\ldots\oplus B_n$ satisfies
$|f|_\xi=|f_1|_\xi+\ldots +|f_n|_\xi$. If for all scalars $|s\bullet
-|_\xi=|s|_\xi\circ|-|_\xi$ then $s, s_1,s_2$ satisfy a distributivity
rule
$s\circ(s_1+s_2)=(s\circ s_1)+(s\circ s_2)$ provided there exists a
morphism
$f:A\to B_1\oplus B_2$ and a scalar
$t$ such that $s_i=|f_i|_\xi$ and $s=|t|_\xi$.  If
$|(-)^\dagger|_\xi=|-|_\xi$ and $|0_{A,B}|_\xi=0_{\II,\II}$ then  all
$f,g\in{\bf C}^\sharp$ satisfy
\beq\label{eq:OBvaluationX} |f\oplus g|_\xi=|f|_\xi+|g|_\xi\,.\quad\ \
\
\eeq
\item[{\bf ii.}]  Let ${\bf C}(\II,\II)\subseteq{\bf C}^\sharp$ and
assume that for all
$s\in|{\bf C}|_\xi$ there exists a scalar
$s^\zeta$ such that
$|s^\zeta|_\xi=s$.  Then for all $s,t\in|{\bf C}|_\xi$ we have
\beq\label{eq:OBmonoidalsum} s+t=|s^\zeta\oplus t^\zeta|_\xi\,,\quad\
\ \ \
\eeq  in the presence of which eq.{\rm(\ref{eq:OBvaluation})} can now
be equivalently rewritten as
\beq\label{eq:OBmonoidal} |f|_\xi=\Bigl|\
|f_1|_\xi^\zeta\oplus|f_2|_\xi^\zeta\,\Bigr|_\xi.\quad\ \ \
\vspace{-3mm}\eeq
\item[{\bf iii.}] If moreover $|-|_\xi:=||-||^{\nu}$ and if the unique
square-roots, the $\nu${\rm th}-powers and
${1\over\nu}${\rm th}-powers consequently required in
eq.{\rm(\ref{eq:OBmonoidalsum})} and eq.{\rm(\ref{eq:OBmonoidal})}
exist,  then eq.{\rm(\ref{eq:OBmonoidal})} rewrites equivalently as
\beq\label{eq:OBmonoidal!} ||f||={\rm
Tr}\bigl(||f_1||\oplus||f_2||\bigr)\,.\quad\ \ \
\eeq
\een
\end{lemma}

By Lemma \ref{superlemma} for $|-|_\xi:=||-||$ and $|-|_\xi:=|-|$ we
respectively have
\[ s +t:={\rm Tr}(s\oplus t)
\ \ \qquad{\rm and}\ \ \qquad s+t=\sqrt{{\rm Tr}\bigl(s^2\!\oplus
t^2\bigr)}=|s\oplus t|
\] assuming self-adjointness of $s$ and $t$. More generally,
$s+t=\bigl({\rm Tr}(s^{1\over\nu}\!\oplus t^{1\over\nu})\bigr)^\nu$
for
$|-|_\xi:=||-||^\nu$ --- the $\nu$th-power outside the trace and the
${1\over\nu}$th-power inside the trace do not cancel out. Different
choices of
$|-|_\xi$ yield different sums and hence different abstract integers
and rationals  e.g.~when setting $2_\II:=1_\II+1_\II$ we have
$2_\II={\rm Tr}(1_{\II\oplus \II})$ for $||-||$ and
$2_\II=\sqrt{{\rm Tr}(1_{\II\oplus \II})}$ for $|-|$ --- recall  here
that
${\rm Tr}(1_{\mathbb{C}\oplus
\mathbb{C}})={\rm dim}(\mathbb{C}\oplus
\mathbb{C})=2$ in {\bf FdHilb}.  The key result of Lemma
\ref{superlemma} is of course eq.(\ref{eq:OBmonoidal!}) which we call
the \em ortho-Bornian axiom\em. We will now decompose this
ortho-Bornian axiom in two tangible components.

\section{Ortho-Bornian structure}\label{sec:OBstructure}

An endomorphism $h:A\to A$ is called \em positive \em iff it
decomposes as
$h=f^\dagger\circ f$. If
$A=A_1\oplus\ldots \oplus A_n$ then we call $h_{11}\oplus \ldots \oplus
h_{nn}:A\to A$ the \em pseudo-diagonal \em  of $h$ with respect to that
decomposition of $A$.  The collection of all positive morphisms of an
SCCC {\bf C} will be denoted by ${\bf C}^+$.  The ortho-Bornian axiom
is now equivalent to validity of
\[ {\rm Tr}(h)={\rm Tr}\bigl({\rm Tr}(h_{11})\oplus{\rm
Tr}(h_{22})\bigr)
\]  for all positive morphisms 
\[
h:= f^\dagger\circ f:A_1\oplus A_2\to A_1\oplus A_2\in {\bf C}^+\,.
\]

\begin{proposition}
In an ortho-SCCC for $h,h'\subseteq{\bf C}^\sharp\cap{\bf C}^+$ we have
\[
{\rm Tr}(h+ h')={\rm Tr}(h\oplus h')
\]
with respect to the sum defined in Theorem \ref{thm:defsum}.  Hence on positive scalars this sum is the one corresponding to the valuation $||-||$  {\rm(}cf.~Lemma \ref{superlemma}\,{\rm)}~i.e.
\[
s +t:={\rm Tr}(s\oplus t)\,.
\]
\end{proposition}
\bpf 
The proof of the first claim proceeds by graphical calculus.  The second claim follows by the fact 
that on scalars ${\rm Tr}(s)=s$.
\endproof

\begin{definition} Let ${\bf C}$ be an ortho-SCCC with ${\bf
C}^+\subseteq {\bf C}^\sharp$.  Its trace satisfies the
\em diagonal axiom \em iff for all $h:A_1\oplus A_2\to A_1\oplus
A_2\in {\bf C}^+$ we have
\[ 
{\rm Tr}(h)={\rm Tr}(h_{11}+ h_{22})
\] and it is \em linear \em iff for all
$h,h'\in {\bf C}^+$ we have
\[ 
{\rm Tr}(h)+{\rm Tr}(h')={\rm Tr}(h+h')\,.
\]
\end{definition}

\smallskip\noindent
%Obviously the diagonal axiom implies for $h_{11},\ldots,h_{nn}\in{\bf C}^\sharp$ that \[ {\rm Tr}(h)={\rm Tr}\bigl(h_{11}+\ldots+ h_{nn}\bigr) \] for all \[ h:A_1\oplus \ldots \oplus A_n\to A_1\oplus \ldots \oplus A_n\in {\bf C}^+\,. \]
Both the diagonal axiom and linearity are stable under the {\it WProj\,}-construction.

\begin{theorem} For an ortho-SCCC {\bf C} with ${\bf C}^+\subseteq{\bf
C}^\sharp$ {\rm TFAE:}
\bit
\item The trace of {\bf C} satisfies the ortho-Bornian axiom.
\item The trace of {\bf C} is linear and satisfies the
diagonal axiom.
\eit
\end{theorem}

The (full) trace ${\rm Tr}(-)=\eta^\dagger_A\circ (1_{A^*}\otimes
-)\circ
\eta_A$ which we have been using so far is a specialization (set
$B=C:=\II$) of the categorical partial trace
\beq\label{Fuckthemall} {\rm
Tr}(-)=\lambda^\dagger_C\circ(\eta^\dagger_A\otimes 1_C)\circ
(1_{A^*}\otimes -)\circ (\eta_A\otimes 1_B)\circ\lambda_B
\eeq which exists as primitive data in so-called \em traced monoidal
categories \em introduced in
\cite{JSV}, and of which compact closed categories are a special
case.  As also shown in
\cite{AC1.5} the required equation for strong compact closure, that is,
eq.(\ref{eq:SCCC}), is equivalent to the \em yanking axiom \em for the
partial trace i.e.
\[
{\rm Tr}(\sigma_{A,A})=1_A\,.
\] 
This allows us to end  with a
conclusive definition in which an ortho-Bornian category arises from
three assumptions on the canonical categorical trace --- the
definition below is not a self-contained definition but relies on the
rest of the paper in order to be understood.

\begin{definition} An \em ortho-Bornian \em SCCC is a category
${\bf C}$ which comes with a special object $\II$ of which the
endomorphisms are called \em scalars\em, with \em tensors \em
$A\otimes B$ and
$f\otimes g$ of objects and morphisms, with \em duals
\em
$A^*$ of objects, with \em adjoints $f^\dagger:B\to A$
\em of morphisms $f:A\to B$, with a special morphism $\eta_A:\II\to A^*\otimes A$
called
\em unit
\em for each object, with \em monoidal sums \em $A\oplus B$ and
$f\oplus g$ of arbitrary objects and of those morphisms which are
included in a subcategory ${\bf C}^\sharp$, all of these pieces of
data being subject to conditions which establish harmonious coexistence
(incl.~Def.~\ref{def:orthostruc}), furthermore ${\bf C}^\sharp$
includes all zero and all positive morphisms, and, the canonical trace
${\rm Tr}(-)$ on ${\bf C}$ which is build from units and their
adjoints as in eq.(\ref{Fuckthemall})
\ben
\item[{\bf 1a.}] satisfies the \em yanking axiom \em as part of the
SCCC-structure,
\item[{\bf 1b.}] satisfies the \em diagonal axiom \em as part of the
ortho-Bornian structure,
\item[{\bf 1c.}] is \em linear \em also as part of the
ortho-Bornian structure.
\een  This category is moreover  \em projective with weights \em iff
it
\ben
\item[{\bf 2.}] satisfies the \em preparation-state agreement axiom\em.
\een
\end{definition}

\section{Weight and relative phase as distinct entities}\label{sec:weight}

Passing from a category such as ${\it WProj}({\bf FdHilb})$ --- or any
other one obtained by applying the {\it WProj\,}-construction to an
SCCC with biproducts --- to a \em genuine
\em ortho-Bornian SCCC involves separating the entities which play the
role of probabilitic weight and of relative phase i.e.~the extra chunk
of state space one gains by considering superpositions of two
underlying state spaces.  In ${\it WProj}({\bf FdHilb})$ these two
entities    respectively are
\[
\mathbb{R}^+:=\{\bar{c}\cdot c\mid c\in\mathbb{C}\}
\quad {\rm and}
\quad
\Bigl\{\bigl\{c\cdot(c_1,c_2)\mid
c\in\mathbb{C}_0\bigr\}\Bigm|(c_1,c_2)\in(\mathbb{C}\times\mathbb{C})_0
\Bigr\}
\] where $\mathbb{C}_0:=\mathbb{C}\setminus \{0\}$ and
$(\mathbb{C}\times\mathbb{C})_0:=\mathbb{C}\times\mathbb{C}\setminus
\bigl\{(0,0)\bigr\}$, hence both are constructed starting from
$\mathbb{C}$, the scalar monoid of
${\bf FdHilb}$. Writing these down when using more conceptual
categorical machinery we get
\[
\bigl\{s^\dagger\!\circ s\bigm| s\in \mathbb{S}\big\}
\qquad\ \   {\rm and}
\qquad\ \
\Bigl\{\bigl\{s\bullet\langle
s_1,s_2\rangle\bigm|s\in\mathbb{S}\bigr\}\Bigm|s_1,s_2\in\mathbb{S}\Bigr\}
\] where $\mathbb{S}:={\bf FdHilb}(\mathbb{C},\mathbb{C})$. The crucial
ingredient which enables us to do this is the pairing operation of the
biproduct structure which allows to express the morphisms
$f\in{\bf FdHilb}(\mathbb{C},\mathbb{C}\oplus\mathbb{C})$ in terms of
those in ${\bf FdHilb}(\mathbb{C},\mathbb{C})$ as $f:=\langle
s_1,s_2\rangle$. But when the ortho-structure of a weighted projective
ortho-Bornian SCCC is not inherited from a biproduct structure we do
not have such a connection. Denoting the scalar monoid as
$\mathbb{W}:={\bf C}(\II,\II)$ --- where every member is now to be
interpreted a probability weight --- the new player is the set
$\mathbb{X}$ implicitly defined within
${\bf C}(\II,\II\oplus\II)=\mathbb{W}\times \mathbb{X}$, that is, the
\em qubit states \em stripped off from any information concerning
probabilistic weight. While these two entities do not share a common
parent anymore they do interact in an important manner via the
\em measurement statistics
\em
\[ {\bf C}(\II,\II\oplus\II)\times{\bf C}(\II,\II\oplus\II)\to
\mathbb{W}::(\psi,\phi)\mapsto
\phi^\dagger\!\circ\psi
\] where ${\bf C}(\II,\II\oplus\II)\times{\bf
C}(\II,\II\oplus\II)\simeq
\mathbb{W}^2\times
\mathbb{X}^2$, and in which the crucial component relating 
probabilistic weight and relative phase
is of type
$\mathbb{X}^2\to\mathbb{W}$.

\end{document}